\documentclass[iop,apjl]{emulateapj}
\usepackage{ulem}

\newcommand{\amm}{NH$_3$}

\newcommand{\kms}{km~s$^{-1}$}

\newcommand{\cmc}{cm$^{-3}$}
\newcommand{\cmq}{cm$^{-2}$}

\newcommand{\ls}{L$_{\odot}$}

\newcommand{\pas}{$\rlap{.}^{\prime\prime}$}

\newcommand{\tkin}{T$_{\rm kin}$}
\newcommand{\trot}{T$_{\rm rot}$}
\newcommand{\ncol}{N$_{\rm col}$}

\slugcomment{\today}

\shorttitle{Hot ammonia in the Orion Hot Core}
\shortauthors{Goddi et al.}

\begin{document}


\title{Unveiling Sources of Heating in the Vicinity of
the Orion BN/KL Hot Core as Traced by Highly Excited Inversion
Transitions of Ammonia}


\author{C. Goddi\altaffilmark{1,2}, L. J. Greenhill\altaffilmark{2}, E. M. L. Humphreys\altaffilmark{1,2}, C. J. Chandler\altaffilmark{3}, and L. D. Matthews\altaffilmark{4}}
\altaffiltext{1}{European Southern Observatory, Karl-Schwarzschild-Strasse 2,
D-85748 Garching bei M$\ddot{u}$nchen}
\altaffiltext{2}{Harvard-Smithsonian Center for Astrophysics, 60 Garden Street, Cambridge, MA 02138}
\altaffiltext{3}{National Radio Astronomy Observatory, P.O. Box O, Socorro, NM 87801}
\altaffiltext{4}{MIT Haystack Observatory, Westford, MA 01886}
\email{cgoddi@eso.org}


\begin{abstract}
Using the Expanded Very Large Array, we have mapped the vicinity  of the Orion BN/KL Hot Core with sub-arcsecond angular resolution in seven metastable inversion transitions of ammonia (NH$_3$): (J,K)=(6,6) to (12,12). This emission comes from levels up to 1500~K above the ground state, enabling identification of source(s) responsible for heating the region.
We used this multi-transition dataset to produce images of the rotational/kinetic temperature (\trot/\tkin) and the column density \ncol~of NH$_3$  for ortho and para species separately and on a position-by-position basis. 
We find \trot~and \ncol~in the range 160$-$490~K and $(1-4) \times 10^{17}$~cm$^{-2}$, respectively. Our spatially-resolved  images show that the highest (column) density and hottest gas is found in a northeast-southwest elongated ridge to the southeast of Source~I.
We have also measured the ortho-para ratio of ammonia, estimated to  vary in the range 0.9-1.6. 
Enhancement of ortho with respect to para and the offset of hot \amm~emission peaks from known (proto)stellar sources provide evidence that the \amm~molecules have been released from dust grains  into the gas-phase through the passage of shocks and not by stellar radiation. 
We propose that the combined effect of Source~I's proper motion and its low-velocity outflow impinging on a pre-existing dense medium  is responsible for the excitation of \amm~and the Orion Hot Core.
Finally, we found for the first time evidence of a slow ($\sim$5~\kms) and compact ($\sim$1000~AU)  outflow towards IRc7. 
\end{abstract}


\section{Introduction}
\label{intro}

Orion BN/KL is the closest high-mass star forming region 
(414$\pm$7~pc; \citealt{Men07}) and a compelling target for studying 
how massive young stellar objects (YSOs) form and interact with their 
surroundings. Objects of considerable long-term interest in BN/KL are
the highly embedded radio Source~I, believed to be a massive YSO \citep{Reid07,Mat10}, 
and the Hot Core $>$1\arcsec~away, which is a rich source of molecular emission \citep[e.g.,][]{Gen89}. 
In dust emission, the compact source SMA1 is visible
2\arcsec-3\arcsec~from Source~I \citep{Beu04}, which also shows rich 
molecular emission
 \citep{Beu05}. On the other hand, 
only weak molecular thermal emission has been detected toward Source~I \citep{Beu05}. However, its surrounding mass flows excite strong H$_2$O and SiO maser emission, permitting investigations of the 3-D gas dynamics at 10-1000~AU radii \citep{Gre04a,Mat10}. 

Despite intensive radio and infrared (IR) study, the primary heating sources for BN/KL ($\sim10^5$~\ls) are still unknown. A strong mid-IR source, IRc2, has been long suspected to be the dominant energy source, 
but when observed at sub-arcsecond resolution, it breaks up into
multiple peaks \citep{Dougados93,Gre04b}. Moreover, mid-IR observations have
recently shown that IRc2 is not self-luminous, but is illuminated and heated by Source~I \citep{Okumura11}. 
While it has been proposed that Source~I is also (externally) heating the Hot Core  \citep{Her88}, direct evidence is lacking and the kinetic temperature (T$_{\rm kin}$) of molecular gas within 1\arcsec~from Source~I is not known. Tackling this uncertainty requires studies at radio rather than IR wavelengths, in order to penetrate the high column densities towards the Hot Core, and  measurements of thermal rather than maser lines. 

 Ammonia (NH$_3$) is well suited to measuring \tkin~over densities $>10^4$~cm$^{-3}$. 
In particular, metastable transitions (J=K) are interesting because they are 
collisionally excited, thus different (optically-thin) transitions can be 
used to determine \tkin~via a rotational diagram analysis \citep{HoTownes83}. 
Inversion transitions from ($J,K$)=(8,8) to (14,14), with upper state energy levels $\sim$800-2000\,K above ground, have been previously detected toward BN/KL with the Effelsberg-100m telescope \citep[30\arcsec\ beam;][]{Wilson93}, confirming the presence of hot molecular gas in the region but not localizing it.  
 \citet{Wilson00} mapped the Hot Core region in the ($J,K$) =(4,4) and (10,9) inversion 
transitions ($\sim$200~K and $\sim$1350~K  above ground, respectively) with $\gtrsim$1\arcsec~resolution. (10,9) is a nonmetastable transition of ortho-\amm~and (4,4) is a metastable transition of para-\amm, hence the Boltzmann analysis required assuming an ortho-para ratio, preventing  an accurate measurement of \tkin. 

In this Letter, we present new observations of (metastable) high-$J$ inversion lines of NH$_3$ in Orion BN/KL. Using the Expanded Very Large Array (EVLA; \citealt{Perley11}), we imaged at sub-arcsecond resolution seven \amm~lines with energy levels high above  the ground state (equivalent to 400-1600~K), from ($J,K$)=(6,6) to (12,12).
The multi-transition measurements enabled estimation of the 
temperature, density, and velocity field of (hot) molecular gas  
and hint at a possible excitation mechanism of the Hot Core. 

\begin{deluxetable*}{cccccccccccc}
\tabletypesize{\tiny}
\tablewidth{0pc}
\tablecaption{Parameters of Observations.}
\tablehead{
\colhead{Transition}\tablenotemark{a} & \colhead{$\nu_{\rm rest}$} & \colhead{$E_u/k$}\tablenotemark{b} & \colhead{Date}  & \colhead{EVLA} & \colhead{Synthesized Beam} & \colhead{RMS}\tablenotemark{c} & \colhead{F$_{\rm peak}$}\tablenotemark{d} & \colhead{F$_{\rm int}$}  & \colhead{V$_{c}$} & \colhead{$\sigma_v$}\\ 
\colhead{(J,K)} & \colhead{(MHz)} & \colhead{(K)} & \colhead{(yyyy/mm/dd)} & \colhead{Receiver}  & \colhead{$\theta_M('') \times \theta_m('')$} &  \colhead{(mJy/beam)}  & \colhead{(Jy)}  & \colhead{(Jy~km/s)}  & \colhead{(km/s)}     & \colhead{(km/s)}
} 
\startdata
(6,6)   & 25056.03    &  408  & 2010/10/11 & K  & $0.93 \times 0.76$ & 3.3 & 2.54 & 30.78 & 5.31 & 4.85 \\
(7,7)   & 25715.18    &  539  & 2010/10/12 & K  & $0.91 \times 0.72$  & 3.3 & 1.62 & 16.25 & 5.14 & 4.01\\
(8,8)   & 26518.91    &  687  & 2010/12/17 & Ka & $0.91 \times 0.79$  & 18. & 1.30 & 10.41 & 4.21 & 3.21\\
(9,9)   & 27477.94    &  853  & 2010/12/21 & Ka & $0.82 \times 0.66$  & 3.1 & 0.88 &  6.89 & 5.37 & 3.15\\
(10,10) & 28604.74    & 1036  & 2010/12/29 & Ka & $0.84 \times 0.65$ & 3.0 & 0.61 &  3.89 & 5.68 & 2.54\\
(11,11) & 29914.49    & 1238  & 2011/01/10 & Ka & $0.77 \times 0.65$  & 2.9 & 0.34 &  1.98 & 5.90 & 2.31\\
(12,12) & 31424.94    & 1456  & 2011/02/05 & Ka & $0.77 \times 0.63$  & 3.5 & 0.17 &  0.82 & 5.93 & 1.91\\
\enddata
\tablecomments{\\
\scriptsize
\tablenotemark{a}{-- Transitions include ortho-\amm~($K=3n$) and para-\amm~($K\neq3n$).} \\
\tablenotemark{b}{--Energy above the ground from the JPL database.} \\ 
\tablenotemark{c}{--RMS noise in a $\sim$0.2~\kms~channel (no velocity-smoothing). 
} \\
\tablenotemark{d}{--F$_{\rm peak}$, F$_{\rm int}$, V$_{c}$, and $\sigma$ are estimated from single-Gaussian fits to the spectral profiles in Figure~\ref{spectra}.}
}
\label{obs}
\end{deluxetable*}
\section{Observations and Data Reduction}
\label{obser}
Observations of NH$_3$ were conducted using the EVLA of the National
Radio Astronomy
Observatory (NRAO).
By using the (new) broadband EVLA K- and Ka-band receivers, we
observed  a total of seven metastable inversion transitions of NH$_3$,
from ($J,K$)=(6,6) to  (12,12)  at 1.3~cm (25-31~GHz).  Transitions
from (6,6) to (11,11) were observed separately in 3h tracks on several
dates using the C configuration.
 The (12,12) dataset was acquired in the hybrid CnB configuration. 
Table~\ref{obs} summarizes the observations.  
Each transition was observed using a 4~MHz bandwidth 
($\sim$40~\kms~at 30~GHz) consisting of 256 channels with a
separation of 15.6~kHz. Typical on-source integration was
1.5h. Each transition was observed with 
``fast switching'', where 80s scans on-target were alternated with 40s
scans of the nearby (1.3$^{\circ}$) QSO J0541$-$0541  (measured flux density $\sim$0.7~Jy). We derived  absolute flux calibration from observations of 3C~48 ($F_{\nu}= 1.1$~Jy) or 3C~147 ($F_{\nu}=$1.4--1.6~Jy), depending on the epoch, and bandpass calibration from observations of 3C~84.

Using the Astronomical Image Processing System (AIPS) task IMAGR, we
 imaged the BN/KL region with cell size 0\pas1, covering a 
50\arcsec~field. 
We   fitted and removed continuum emission from spectral cubes 
using AIPS' IMLIN.
  The data were processed both with and without velocity-smoothing, 
resulting in velocity resolutions of 0.15--0.19~\kms\ and 0.6--0.8~\kms,
respectively  (depending on transition).
The following analysis is based on the smoothed dataset.
To match the angular resolution of different configurations, we
  produced maps 
  setting the ``ROBUST'' weighting parameter to ${\cal R}$=0 for the transitions
 observed with the C-array and to ${\cal R}$=5 (natural weighting) 
for the (12,12) transition observed with the CnB-array.  
We restored all images with a beam of size 0\pas85$\times$0\pas7 (P.A.=0),
approximately the average size among different transitions (see
 Table~\ref{obs}). After smoothing, the RMS noise per
 channel was typically $\sim$1.6~mJy~beam$^{-1}$.
The (8,8) transition falls at the band-edge of the Ka-band receiver
 where performance was not optimal, resulting in lower-quality data
 compared with the other transitions. Hence, we excluded it from the
 following analysis. 
Moreover, inspection of data in post-processing revealed a 30\% offset  in the absolute flux scale for the (9,9) dataset, possibly due to instability of the Ka-band system during the observations.
We accounted for the systematic error by scaling the flux density by 30\% in the final image of the (9,9) transition.

\begin{figure}
\centering
\includegraphics[width=0.5\textwidth]{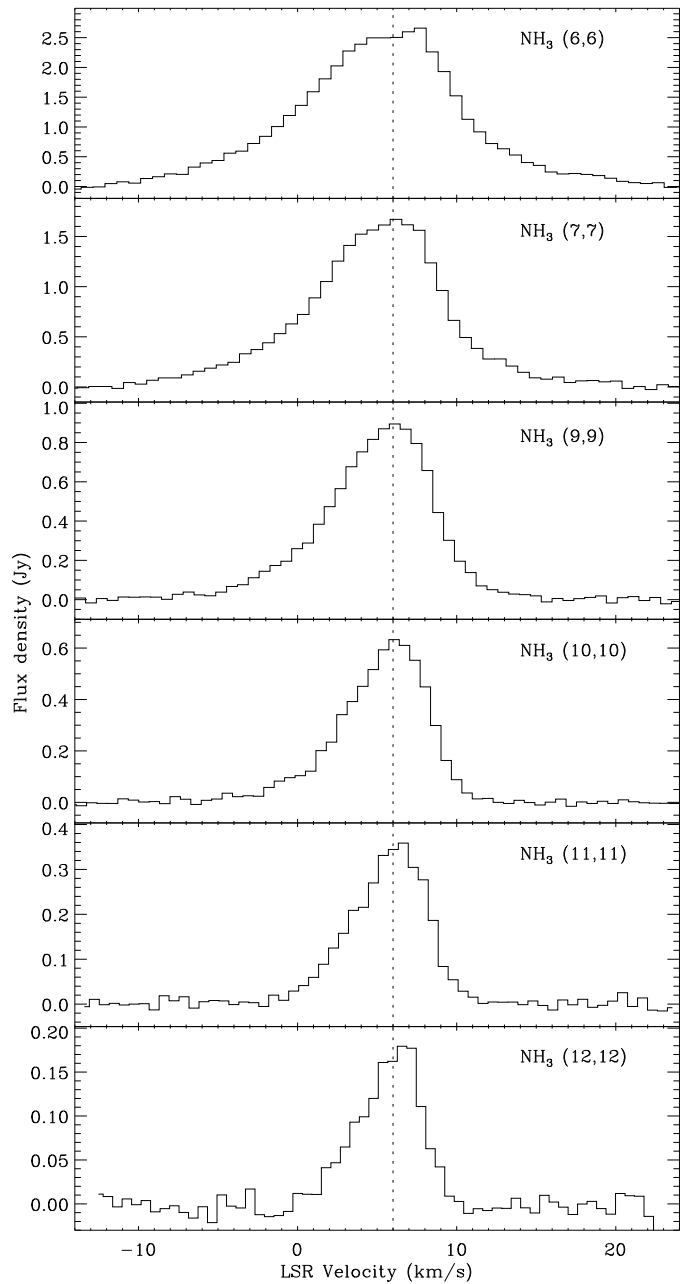}
\caption{Spectra of various NH$_3$ inversion transitions observed toward Orion~BN/KL with the EVLA. 
The velocity resolution is 0.7~\kms. The flux density is integrated over the Hot Core region and the radial velocities are with respect to the local standard rest (LSR). The vertical dashed line indicates a velocity of 6~\kms. Note that different flux density scales are adopted for different transitions.
}
\label{spectra}
\end{figure}

\begin{figure*}
\centering
\includegraphics[width=0.65\textwidth,angle=-90]{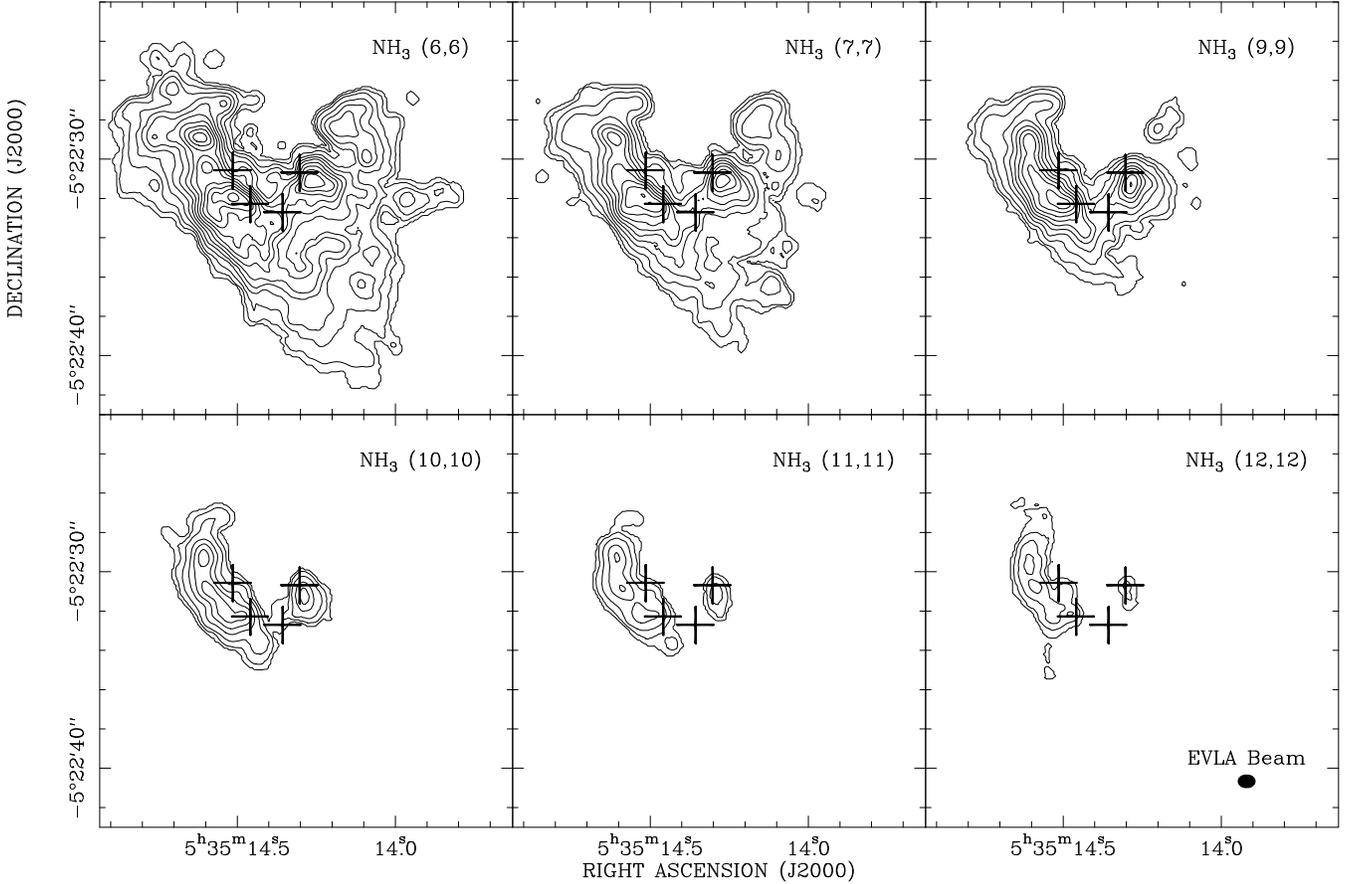}
\caption{Total intensity images of NH$_3$ inversion transitions  from (6,6) to (12,12) as observed toward Orion~BN/KL with the EVLA. 
The images were integrated over the velocity range 
--14~\kms~to 24~\kms\ with a flux cutoff of 5~mJy~beam$^{-1}$
($\sim3\sigma$) for all transitions. 
Contours are 1, 20, 50,.., 550 (by 50) mJy~beam$^{-1}$~\kms. 
The beam (0\pas85$ \times$ 0\pas7) is shown in the right-lower corner. 
In every panel, the crosses mark the position of 4 YSOs in the region: 
sources~I, SMA1, {\it n}, and IRc7 (left to right).
}
\label{mom0}
\end{figure*}

\section{Results}

For the first time we have mapped and localized the hot \amm~gas from metastable transitions (6,6) up to (12,12) with sub-arcsecond resolution towards BN/KL.  
We observed three ortho (6,6; 9,9; 12,12) and three para (7,7; 10,10; 11,11) transitions so as to bracket an upper-state energy
of 1000\,K without exceeding 2000\,K.

For each transition, we produced spectra by mapping each spectral channel and summing the flux density in each channel map (Figure~\ref{spectra}). Multiple transitions show similar line profiles and central velocities ($\sim$5-6~\kms), while velocity widths, $\sigma_v$, systematically decrease from (6,6)$\sim$5~\kms~to (12,12)$\sim$2~\kms, as determined from single-Gaussian fits (Table~\ref{obs}).
By comparing with spectral profiles from the Effelsberg-100m telescope (assuming an equivalence relation K=0.86~Jy; \citealt{Wilson93}), most of the single-dish flux density for transitions from (9,9) to (12,12) was recovered by the  EVLA. This is not surprising considering that these transitions arise from levels $>$850~K above the ground, and are thus expected to arise from compact regions close to the exciting source.


We produced total intensity images of various NH$_3$ inversion
transitions, integrated over the whole band (Fig.~\ref{mom0}). 
In the image we included also the positions of the four YSOs Source~I, {\it n}, SMA1, and IRc7.
The low-excitation \amm~emission (6,6; 7,7) is extended over
$\sim15^{\prime \prime }\times 15^{\prime \prime }$ ($0.03 \times
0.03$~pc) and shows the typical Hot Core ``heart-shaped'' structure
previously observed in various 
molecular lines:  C$_2$H$_5$CN, CH$_3$OH, OCS  \citep[]{FriedelSnyder08}, CH$_3$CN \citep[]{Zapata11b}, and \amm~(4,4) \citep[]{Wilson00}. 
The northwest (NW) lobe has the weakest integrated emission and it is not detected in the higher-$J$ transitions,  
whose emission originates from the northeast lobe of the heart-shaped structure. 
This confirms earlier results from \citet{Wilson00} based on a comparison between (4,4) and (10,9). Interestingly, the highly-excited \amm~line emission shows an ``arc-like'' ridge, oriented northeast-southwest (NE-SW), $\lesssim$3000~AU long and $\lesssim$1000~AU across, curving around Source~I. Source~I lies at the NW edge, indicating a lower column density towards its position. 

West of Source~I, we find high-$J$ emission associated with IRc7. The structure of this emission in the transitions from (10,10) to (12,12) is bipolar and elongated north-south (N-S), with size $\sim$1000~AU.

\subsection{Temperature/density analysis}
\label{rot_dia}
We estimated rotational temperature, \trot, and column density, \ncol, from rotational energy diagrams, using six observed  transitions.  We assumed that the inversion transitions are thermalized and optically thin. The beam-averaged \ncol\ from an optically thin transition is given by \citep{FriedelSnyder08}:
\begin{equation}
\label{ntot}
 \frac{N_{col}}{Q} e^{-E_{u}/T_{rot}} = \frac{2.04 W}{ \theta_M \theta_m S\mu^2\nu^3} \times 10^{20} \rm cm^{-2}  
\end{equation} 
 where $Q$ is the partition function, $E_u$ is the upper state energy of the transition (K), $W$ is the integrated line intensity (Jy~beam$^{-1}$~\kms), $\theta_M$ and $\theta_m$ are the FWHM Gaussian synthesized beam dimensions (arcseconds), $S\mu^2$ is the product of the line strength and the square of the molecular dipole moment (Debye$^2$), and $\nu$ is the transition frequency (GHz).
For H$_2$ densities in the Hot Core ($\sim 10^7$~\cmc; \citealt{Gen89}) we assume \trot=\tkin~\citep{Wilson00}, so in the following we simply refer to \tkin.

We performed separate temperature analyses for ortho and para
species, enabling empirical determination of the \amm~ortho-para
ratio in the Hot Core: 0.9-1.6 ($\pm0.05$ - 0.1) with an average
over all pixels of 1.1. The highest values are located
in the NW-side of the ridge which faces Source~I.  

\begin{figure*}
\centering
\includegraphics[width=0.55\textwidth]{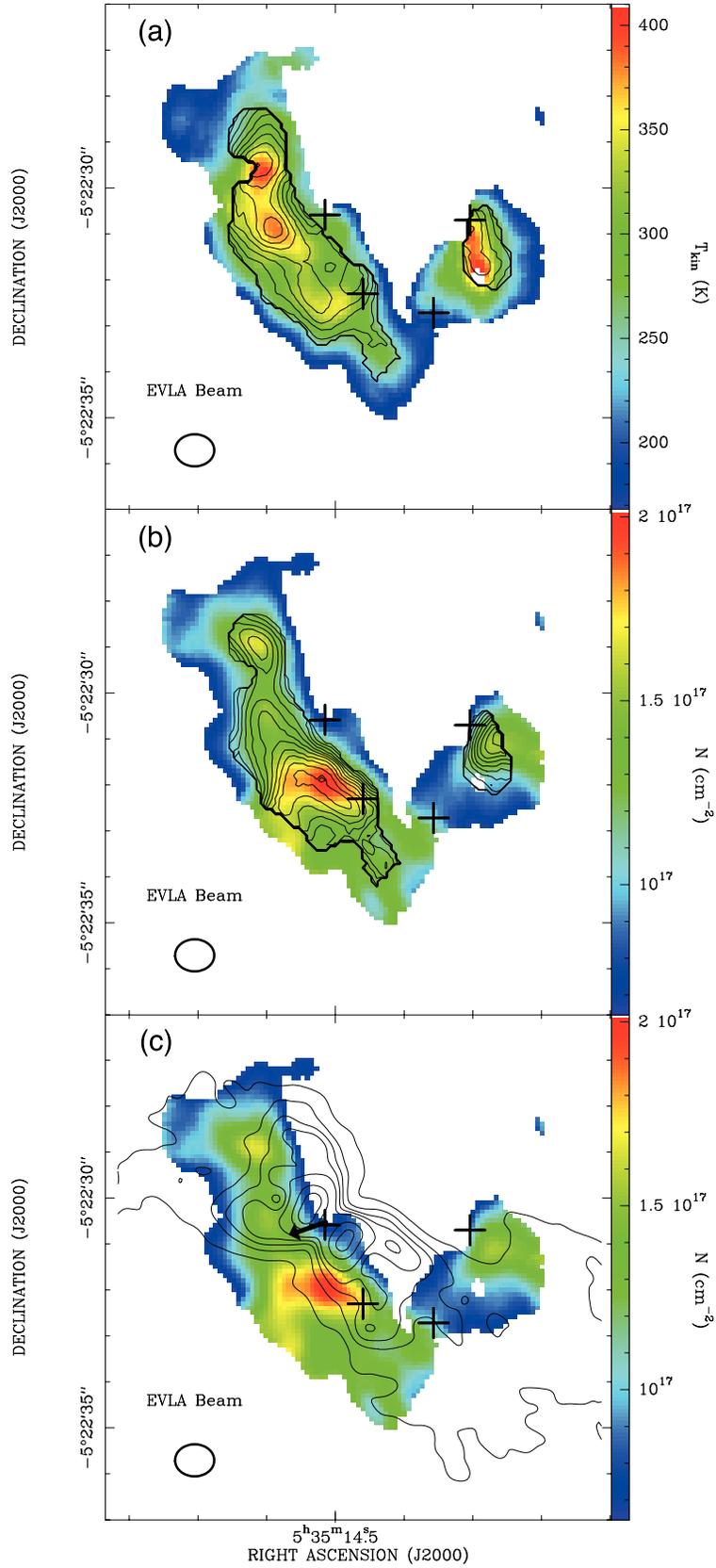}
\caption{\scriptsize \amm\ rotational temperature (\trot) and column density (\ncol)
maps of Orion~BN/KL, obtained from ortho ({\it colors}) and para 
({\it contours} in top two panels) inversion transitions. In all
panels, the crosses mark the positions of sources I, SMA1, {\it n}, and IRc7 
(left to right).
(a) Map of \trot~averaged over LSR 
velocities --14~\kms~to 24~\kms. 
Contour levels are 210~K to 480~K in increments of 30~K. Outside the
  region 
shown, the signal-to-noise ratio was insufficient to
  calculate 
\trot~($\Delta T>$8~K) reliably. 
(b) Map of \ncol~obtained using equation~1. Contours are from 
0.8 to 2 $\times 10^{17}$~cm$^{-2}$ in increments of
  $10^{16}$~cm$^{-2}$  
($\Delta N \lesssim 10^{16}$~\cmq). 
Note that the densest and hottest \amm~are found in a 
northeast-southwest elongated ridge to the southeast of Source~I.
(c) Ortho \ncol~({\it colors}) with SiO $v$=0, $J=2-1$ total intensity  
({\it contours}) from \citet{Plambeck09} overplotted. Contours
  are 0.1, 0.3, 0.5, 0.7, 1, 2, 3 Jy~beam$^{-1}$~\kms. The arrow
  indicates the proper motion of Source~I in the Orion rest-frame \citep{Goddi11}. 
}
\label{tkin_ncol}
\end{figure*}

We determined the distribution of \tkin~and \ncol~on a position-by-position basis. We measured \tkin$\sim$160-490~K ($\Delta T\lesssim8$~K) and \ncol$\sim (1-4) \times 10^{17}$~\cmq  ($\Delta N \lesssim 10^{16}$~\cmq), averaged on scales $\lesssim$1\arcsec (Fig.~\ref{tkin_ncol}). 
 The highest values of \tkin~and \ncol 
 lie in a NE-SW elongated ridge offset (to the southeast) from Source~I.
At Source~I's position, we measure \tkin$\sim$260~K and \ncol$\sim 7\times 10^{16} $~\cmq. Temperature and density towards Source~I are estimated using only ortho transitions because the signal-to-noise ratios in para data were insufficient. Consequently, total column density is probably 1.5-$2\times$ greater.

\citet{Wilson93} reported $400\pm40$\,K using the inversion lines
(10,10) to (14,14) observed at 30\arcsec~resolution, while
\citet{Wilson00} reported values in the range 130-170~K from
1\arcsec~images of (4,4) and (10,9). 
In the paper presented here, high-excitation transitions observed at
sub-arcsecond resolution enable us to map the gas density and
temperature distribution with unprecedented precision.
This has revealed hotter, denser material than previously derived from \amm~measurements. The temperature inferred from high-$J$ lines of CH$_3$CN is 200-600\,K \citep{Wang10}, consistent with our findings from \amm. 

Our analysis assumes optically thin transitions \citep{Wilson00}. Hyperfine-structure satellite lines are outside of the bandwidths employed, hence direct estimation of optical depth is impossible. \citet{Her88} reported optical depths $>$1 for the (6,6) and (7,7) transitions. Non-negligible optical depths in the lowest available transitions (6,6 for ortho; 7,7 for para) might systematically boost the derived temperatures. We argue otherwise that this is not the case.
Para-transitions yield higher temperatures than ortho-transitions, inconsistent with expectation of lower optical depth for (7,7) than (6,6). 
Moreover,  rotation diagrams for individual pixels do not diverge from a linear trend for the lower (J,K) states.

\subsection{Velocity field}
\begin{figure*}
\centering
\includegraphics[width=0.6\textwidth,angle=-90]{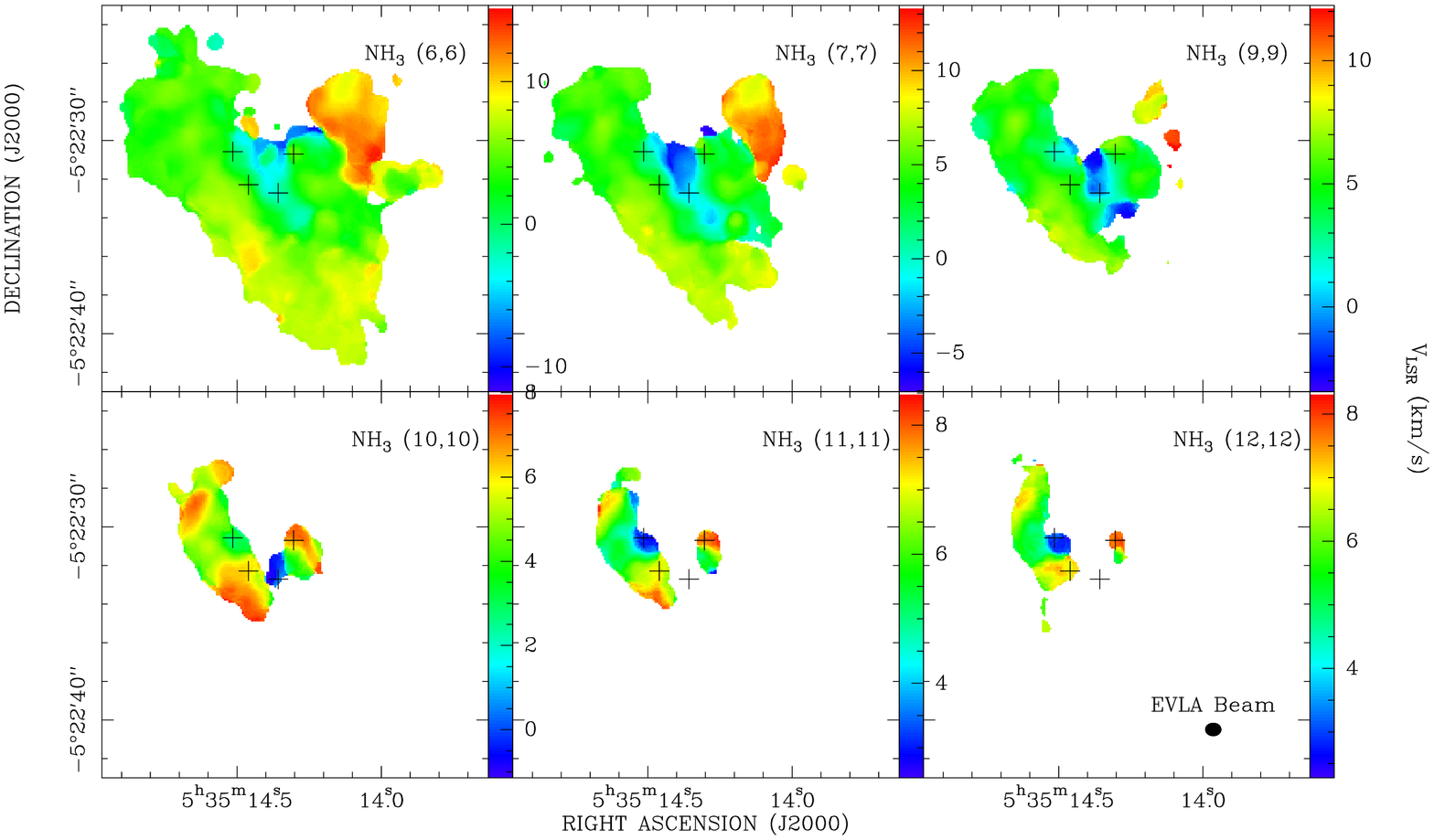}
\caption{Velocity fields of \amm~inversion transitions from (6,6) to (12,12). Colors indicate V$_{LSR}$ in \kms. 
Color scales are compressed with increasing quantum number to clearly show velocity structure. In each panel, the crosses mark the position of sources I, SMA1, {\it n}, and IRc7 (from left to right). 
All transitions show a velocity gradient southward from Source~I towards the Hot Core center (the less clear signature in the upper panels is due to the expanded color scale). 
}
\label{mom12}
\end{figure*}

The first moment maps of (6,6) and (7,7) in Figure~\ref{mom12} 
show that redshifted gas (7 to 14~\kms) is located on the NW part of the heart-shaped structure, blueshifted gas ($-$10 to 0~\kms) is concentrated towards the central part, 
while the NE lobe is around systemic velocity (0 to 7~\kms).  
The \amm~emission shows a velocity gradient southward from Source~I towards the Hot Core center, where the largest values of velocity dispersion are also observed. 
Unless we assume the presence of an additional embedded source in that location, this may indicate shock propagation in the Hot Core (see Sect.~\ref{discussion}). 
Interestingly, we identified a N-S velocity gradient  ($\Delta v
\sim$5~\kms) in the bipolar structure traced by high-$J$ \amm~emission
in IRc7, where line-widths also appears enhanced. The N-S bipolar morphology, the velocity gradient in the elongation axis, and the line-width enlargement provide strong evidence of a low-velocity outflow associated with IRc7.  


\section{Discussion \& Conclusions}
\label{discussion}
 
Our multi-transition measurements of  \amm~emission enable the estimation of temperature  and density of (hot) molecular gas at high-spatial resolution in the vicinity of Source~I and the Hot Core. 
This complements the temperature distribution as a function of radius close to Source~I inferred from radiative transfer modelling of SiO masing gas \citep{Goddi09}.  Maser action in the disk/wind from Source~I implies temperatures  1000-2000\,K inside $\sim$100\,AU and  temperatures of 400-1000\,K  in the outflow at radii 100~AU to 1000~AU. 
A dust color temperature of $\sim$700~K is consistent with this finding \citep{Okumura11}.
On the other hand, an \amm~temperature of $\sim$260~K toward Source~I implies that hot NH$_3$ emission is not directly (radiatively-)excited by Source~I. 
One explanation of this apparent discrepancy could be that a high ultraviolet radiation field in the vicinity of Source~I may dissociate fragile molecules like \amm~(dissociation energy 4~eV), while enhancing gas-phase abundance of SiO through grain sputtering. This would also explain the offset between Source~I and the peak of high-$J$ \amm~emission. 

What is exciting the high-energy transitions of \amm?
In fact, all known or suspected (massive) YSOs in the region are offset from peaks in hot \amm~emission (Figure~\ref{mom0}), and there is no evidence of any other embedded objects heating the Hot Core in the radio \citep[]{MentenReid95}, millimeter \citep[]{FriedelSnyder08}, and infrared \citep[]{Gre04b}. 
Additionally, the color temperature distribution from mid-IR measurements shows a decreasing gradient from Source~I to the Hot Core \citep{Okumura11}.
We propose that the elevated gas temperatures measured from \amm~in the Hot Core are produced by shock-waves  and not by stellar radiation. 
Theoretical models suggest that \amm~might form through surface chemistry on dust grains and then be released  in the gas-phase by the passage of shocks, either through hydrodynamic gas-grain interaction \citep{Flower94,Flower95} or shock-induced IR radiation \citep{Taylor96}. 
Indeed, enhanced \amm~has been observed in the L1157 low-mass outflow \citep{Tafalla95}, as well as towards a dense (quiescent) clump downstream from the high-mass protostellar jet HH 80 North \citep{Girart98}.
\citet{Umemoto99} report also ortho-para enhancement in the L1157 outflow.
Several pieces of evidence corroborate the shock hypothesis in the Orion Hot Core: the offset of hot \amm~emission peaks from known (proto)stellar sources, the high \tkin, the departure of the ortho-para ratio from the statistical equilibrium value (1.0; \citealt{Umemoto99}), and the high abundances of shock tracer molecules (SiO, SO, SO$_2$) and common molecular constituents of interstellar ices  (C$_2$H$_5$CN, CH$_3$OH, CH$_3$CN; \citealt{Vandishoeck98}).

What generates the shocks? 
Recent studies report strong evidence that a dynamical interaction
occurred $\sim$500~years ago between Source~I and the  high-mass
YSO BN  \citep{Gomez08,Goddi11}. The stellar interaction resulted in 
high stellar proper motions for  Source~I  and BN and possibly
produced the fast bullet outflow traced by  CO and H$_2$ 
\citep{Zapata09,Bally11,Goddi11}.  
In this framework, it has recently been proposed that the Hot Core
might have originated from the impact of a shock-wave onto a
pre-existing dense core, driven by either the expansion of a dense
bubble of material previously associated with BN and Source~I 
\citep{Zapata11a} and/or the  fast bullets \citep{Zapata11b}. 
Data in Figure~\ref{tkin_ncol}c, however, suggest another possibility. 
The relationship between the high column density gas as traced by
\amm,  the Source~I proper motion, and the Source~I outflow as traced by SiO emission \citep{Plambeck09} 
together suggest that the Source~I outflow drives compression and shocks in the surrounding medium.
 The relative narrowness of the SW outflow
in close proximity of the column density peak is suggestive of ram pressure
effects caused by the outflow and the YSO proper motion.
The more open NE lobe, its apparent symmetry about the flow axis, and its
extension beyond the northern NH$_3$ clump may indicate that the outflow is
in front of or behind the dense gas, compressing it more along the line of sight.
Part of the mechanical energy of the outflow could be dissipated and 
contribute to gas heating. This is also consistent 
with the ortho-para ratio being larger in the NW-side of the ridge
facing Source~I.

Detailed modeling, including shock-physics as well as dust/gas-phase
chemistry, will be required to discriminate between different (shock)
scenarios proposed for the Hot Core excitation.
However, the shock velocities implied by different scenarios and inferred \amm~chemistry are suggestive. \citet{Flower94} showed that shocks with velocities of 10--50~\kms~(C-type) efficiently sublimate ice mantles, ejecting volatile molecules intact and enabling efficient abundance enhancement.
Shocks with velocities $<$10~\kms~would have just a diminished impact on grain-mantle evaporation and  \amm~abundance enhancement, which weighs against the hypothesized slowly-expanding bubble (7~\kms)\footnote{The 15~\kms~value quoted in \citet{Zapata11a} is divided by two for spherical expansion.} as the main mechanism for the Hot Core excitation.
In contrast, fast bullets ($>100$~\kms) would drive strong J-type shocks, causing grain sputtering and molecular dissociation. Reformation of \amm~by gas-phase chemistry is possible but less likely on the short dynamical time of the region ($\sim$500~years).  Nonetheless, existing models have not investigated chemical evolution behind J-type shocks. Intermediate-velocity bullets traced by H$_2$ (30-70~\kms; \citealt{Bally11}) could in principle drive C-type shocks and release \amm~as well as excite H$_2$ emission, but the poor correlation between \amm~and mid-IR/H$_2$ emission \citep{Shu04} argues against this possibility.
We note that the impact of outflow from Source~I (20~\kms) in combination with proper motion of 12~\kms could efficiently heat the Hot Core with limited molecular dissociation.  
\citet{Favre11} reported an anticorrelation of methyl-formate and H$_2$ in the Hot Core and proposed a similar mechanism for its excitation. Extinction could however be responsible for the observed anticorrelation.

In summary, we suggest that local mechanical interaction of Source~I with the Orion Hot Core is more likely to be responsible for enhanced \amm~abundance and heating of gas in the Core than the consequence of the hypothesized explosive event tied to the protostellar dynamical interaction.

\acknowledgments{
We thank Malcolm Walmsley and Steve Longmore for useful discussions, and the anonymous referee for a constructive report.
These data were obtained under EVLA OSRO program 10B-225 and
the work was supported by the National Science Foundation (AST 0507478).}



\begin{thebibliography}{}
\bibitem[Bally et al.(2011)]{Bally11} Bally, J., et al.\ 2011, \apj, 727, 113 
\bibitem[Beuther et al.(2004)]{Beu04} Beuther, H., et al.\ 2004, \apjl, 616, L31 
\bibitem[Beuther et al.(2005)]{Beu05} Beuther, H., et al.\ 2005, \apj, 632, 355
\bibitem[Dougados et al.(1993)]{Dougados93} Dougados, C., Lena, P., Ridgway, S.~T., Christou, J.~C., \& Probst, R.~G.\ 1993, \apj, 406, 112 
\bibitem[Favre et al.(2011)]{Favre11} Favre, C., et al.\ 2011, arXiv:1103.2548 
\bibitem[Flower 
\& Pineau des Forets(1994)]{Flower94} Flower, D.~R., \& Pineau des Forets, G.\ 1994, \mnras, 268, 724 
\bibitem[Flower et 
al.(1995)]{Flower95} Flower, D.~R., Pineau des Forets, G., \& Walmsley, C.~M.\ 1995, \aap, 294, 815 
\bibitem[Friedel \& Snyder(2008)]{FriedelSnyder08} Friedel, D.~N., \& Snyder, L.~E.\ 2008, \apj, 672, 962
\bibitem[Genzel \& Stutzki(1989)]{Gen89} Genzel, R., \& Stutzki, J.\ 1989, \araa, 27, 41 
\bibitem[Girart et al.(1998)]{Girart98} Girart, J., Estalella, 
R., \& Ho, P.~T.~P.\ 1998, \apjl, 495, L59 
\bibitem[Goddi et al.(2009)]{Goddi09} Goddi, C., Greenhill, 
L.~J., Chandler, C.~J., Humphreys, E.~M.~L., Matthews, L.~D., 
\& Gray, M.~D.\ 2009, \apj, 698, 1165 
\bibitem[Goddi et al.(2011)]{Goddi11} Goddi, C., Humphreys, 
E.~M.~L., Greenhill, L.~J., Chandler, C.~J., \& Matthews, L.~D.\ 2011, \apj, 728, 15 
\bibitem[G{\'o}mez et al.(2008)]{Gomez08} G{\'o}mez, L., 
Rodr{\'{\i}}guez, L.~F., Loinard, L., Lizano, S., Allen, C., Poveda, A., 
\& Menten, K.~M.\ 2008, \apj, 685, 333 
\bibitem[Greenhill et al.(2004a)]{Gre04a} Greenhill, L.~J., 
Reid, M.~J., Chandler, C.~J., Diamond, P.~J., 
\& Elitzur, M.\ 2004a, Star Formation at high-angular Resolution, 221, 155
\bibitem[Greenhill et al.(2004b)]{Gre04b} Greenhill, L.~J., 
Gezari, D.~Y., Danchi, W.~C., Najita, J., Monnier, J.~D., 
\& Tuthill, P.~G.\ 2004b, \apjl, 605, L57 
\bibitem[Hermsen et al.(1988)]{Her88} Hermsen, W., Wilson, T.~L., Walmsley, C.~M., \& Henkel, C.\ 1988, \aap, 201, 285 
\bibitem[Ho \& Townes(1983)]{HoTownes83} Ho, P.~T.~P., \& Townes, C.~H.\ 1983, \araa, 21, 239 
\bibitem[Matthews et al.(2010)]{Mat10} Matthews, L.~D., 
Greenhill, L.~J., Goddi, C., Chandler, C.~J., Humphreys, E.~M.~L., 
\& Kunz, M.~W.\ 2010, \apj, 708, 80 
\bibitem[Menten \& Reid(1995)]{MentenReid95} Menten, K.~M., \& Reid, M.~J.\ 1995, \apjl, 445, L157 
\bibitem[Menten et al.(2007)]{Men07} Menten, K.~M., Reid, M.~J., Forbrich, J., \& Brunthaler, A.\ 2007, \aap, 474, 515
\bibitem[Okumura et al.(2011)]{Okumura11} Okumura, S.-i., 
Yamashita, T., Sako, S., Miyata, T., Honda, M., Kataza, H., 
\& Okamoto, Y.~K.\ 2011, arXiv:1104.4394 
\bibitem[Perley et al.(2011)]{Perley11} Perley, R.A., Chandler, C.J., Butler, B.J., Wrobel, J.M.\ 2011, \apjl, in press 
\bibitem[Plambeck et al.(2009)]{Plambeck09} Plambeck, R.~L., et 
al.\ 2009, \apjl, 704, L25 
\bibitem[Reid et al.(2007)]{Reid07} Reid, M.~J., Menten, K.~M., Greenhill, L.~J., \& Chandler, C.~J.\ 2007, \apj, 664, 950
\bibitem[Shuping et al.(2004)]{Shu04} Shuping, R.~Y., Morris, M., \& Bally, J.\ 2004, \aj, 128, 363 
\bibitem[Tafalla \& Bachiller(1995)]{Tafalla95} Tafalla, M., \& Bachiller, R.\ 1995, \apjl, 443, L37 
\bibitem[Taylor 
\& Williams(1996)]{Taylor96} Taylor, S.~D., \& Williams, D.~A.\ 1996, \mnras, 282, 1343 
\bibitem[Umemoto et al.(1999)]{Umemoto99} Umemoto, T., Mikami, 
H., Yamamoto, S., \& Hirano, N.\ 1999, \apjl, 525, L105 
\bibitem[van Dishoeck \& Blake(1998)]{Vandishoeck98} van Dishoeck, E.~F., \& Blake, G.~A.\ 1998, \araa, 36, 317 
\bibitem[Wang et al.(2010)]{Wang10} Wang, K.-S., Kuan, Y.-J., 
Liu, S.-Y., \& Charnley, S.~B.\ 2010, \apj, 713, 1192 
\bibitem[Wilson et al.(1993)]{Wilson93} Wilson, T.~L., Henkel, C., Huttemeister, S., Dahmen, G., Linhart, A., Lemme, C., \& Schmid-Burgk, J.\ 1993, \aap, 276, L29 
\bibitem[Wilson et al.(2000)]{Wilson00} Wilson, T.~L., Gaume, 
R.~A., Gensheimer, P., \& Johnston, K.~J.\ 2000, \apj, 538, 665 
\bibitem[Zapata et al.(2009)]{Zapata09} Zapata, L.~A., 
Schmid-Burgk, J., Ho, P.~T.~P., Rodr{\'{\i}}guez, L.~F., 
\& Menten, K.~M.\ 2009, \apjl, 704, L45 
\bibitem[Zapata et al.(2011a)]{Zapata11a} Zapata, L.~A., Loinard, 
L., Schmid-Burgk, J., Rodr{\'{\i}}guez, L.~F., Ho, P.~T.~P., 
\& Patel, N.~A.\ 2011a, \apjl, 726, L12 
\bibitem[Zapata et 
al.(2011b)]{Zapata11b} Zapata, L.~A., Schmid-Burgk, J., \& Menten, K.~M.\ 2011b, \aap, 529, A24 



\end{thebibliography}
\end{document}